\documentclass[a4paper,11pt]{article}
\pdfoutput=1 
\usepackage{jheppub}
\usepackage[T1]{fontenc}
\usepackage{stmaryrd}
\usepackage{amssymb,amsmath,amsfonts,nccmath,amsthm}
\usepackage{hyperref}
\usepackage{IEEEtrantools}
\usepackage{float}
\usepackage{physics}
\usepackage{braket}
\usepackage{tensor}
\usepackage{simplewick}
\usepackage{graphicx}
\usepackage{xcolor}
\usepackage{tensor}
\usepackage[utf8]{inputenc}
\usepackage{caption,subfig}
\usepackage{bm,bbm,bbold}	
\usepackage{tikz}	
\usepackage{scalerel}
\usepackage{feynmp-auto}
\usepackage{booktabs}
\usepackage{longtable}
\usepackage{verbatim}
\usepackage{eurosym}
\usepackage{enumitem}

\usetikzlibrary{arrows,positioning,decorations.markings,decorations.pathmorphing,calc}

\newcommand{\E}{}
\def\E#1#{\tensor#1{\epsilon}}

\newcommand{\pd}{\partial}

\def \be {\begin{equation}}
\def \ee {\end{equation}}

\title{Quantum stability of Proca-Nuevo}
\author[a]{Claudia de Rham,}
\author[b]{Lavinia Heisenberg,}
\author[c]{Ankip Kumar}
\author[b]{and Jann Zosso}
\affiliation[a]{Theoretical Physics, Blackett Laboratory, Imperial College, London, SW7 2AZ, U.K.}
\affiliation[b]{Institute for Theoretical Physics,
ETH Z\"urich, Wolfgang-Pauli-Strasse 27, 8093, Z\"urich, Switzerland}
\affiliation[c]{Arnold Sommerfeld Center for Theoretical Physics, Department f\"ur Physik, Ludwig-Maximilians-Universi\"at M\"unchen, Theresienstr. 37, 80333 M\"unchen Germany}
\emailAdd{c.de-rham@imperial.ac.uk}
\emailAdd{lavinia.heisenberg@phys.ethz.ch}
\emailAdd{ankipkumar11@gmail.com}
\emailAdd{jzosso@phys.ethz.ch}
\abstract{
 The construction of general derivative self-interactions for a massive Proca field relies on the well-known condition for constrained systems of having a degenerate Hessian. The nature of the existing constraints algebra will distinguish among different classes of interactions. Proca-Nuevo interactions enjoy a non–trivial constraint by mixing terms of various order whereas Generalized Proca interactions satisfy the degeneracy condition order by order for each individual Lagrangians. In both cases the vector field propagates at most three degrees of freedom. It has been shown that the scattering amplitudes of Proca-Nuevo arising at the tree level always differ from those of the Generalized Proca, implying their genuinely different nature and a lack of relation by local field redefinitions. In this work, we show the quantum stability of the Proca-Nuevo theory below a specific UV cut-off. Although Proca-Nuevo and Generalized Proca are different inherently in their classical structure, both have the same high energy behaviour when quantum corrections are taken into account. The arising counter terms have the exact same structure and scaling. This might indicate that whatever UV completion they may come from, we expect it to be of similar nature.

}

\keywords{Proca theories, quantum field theory, effective field theory, dark energy}
\begin{document}
	\allowdisplaybreaks[1]
	\maketitle
	\flushbottom
	%

\section{Introduction}\label{Sec:Introduction}


Making sense of the fundamental quantum nature of matter withing the context of General Relativity (GR) \cite{Einstein_1916} is the holy grail of modern theoretical physics. Intriguingly, discrepancies do not restrict to the UV where ignorance is easily acknowledged, but are already strikingly present in the IR picture of gravity, where according to a Wilsonian effective field theory (EFT) point of view everything should be well understood. The most prominent example is the cosmological constant problem \cite{RevModPhys.61.1} which has gained a new twist with the evidence for an accelerating expansion of the universe \cite{Riess:1998cb,Perlmutter:1998np}, providing the main motivation for a multitude of proposals of low energy modifications of Einstein gravity \cite{Heisenberg:2018vsk}.

A major novelty was put forward by the Dvali-Gabadadze-Porrati (DGP) braneworld model \cite{Dvali:2000hr}, a five-dimensional model with effectively massive gravitons in $4d$, which incorporates a self accelerating branch as cosmological solution as an alternative to a finite cosmological constant \cite{Einstein_1917}. Despite the fact that the self accelerating branch suffers from a ghost instability \cite{Luty:2003vm,Nicolis:2004qq,Koyama:2005tx} the idea inspired a wide rage of research. In particular, the decoupling limit on the brane of DGP is governed by a cubic higher derivative self-interaction of the helicity-0 mode which naturally incorporates a Vainshtein screening mechanism \cite{VAINSHTEIN,Deffayet:2001uk,Kimura:2011dc,Babichev:2013usa,Koyama:2013paa,Kase:2013uja} while evading an Ostrogradsky instability \cite{Ostrogradsky:1850fid,Woodard:2015zca}. Soon it was realized, that there exists a finite set of higher order scalar interactions with the same properties, the Galileons \cite{Nicolis:2008in}, leading to a rediscovery of the covariant Horndeski theory \cite{Horndeski:1974wa} as well as a formulation of a vector counter part known as Generalized Proca (GP) \cite{Heisenberg:2014rta,Allys:2015sht,Jimenez:2016isa} with various cosmological applications on their own right. A crucial aspect of effective field theories (EFT) of this type is their radiadive stability in regimes where the classical irrelevant interactions dominate such that the theory remains viable on scales relevant for the Vainshtein screening, in other words that there exists a regime for which classical non-linearities dominate, while quantum effects are still under control \cite{Luty:2003vm,Nicolis:2004qq,Burgess:2006bm,Hinterbichler:2010xn,Goon:2016ihr,dePaulaNetto:2012hm,Brouzakis:2013lla,Brouzakis:2014bwa,Pirtskhalava:2015nla,Heisenberg:2019udf,Heisenberg:2019wjv,Heisenberg:2020cyi,Heisenberg:2020jtr}.

Recently, a new class of Proca interactions was introduced \cite{ProcaNeuvo} dubbed ``Proca-Nuevo'' (PN). As opposed to Generalized Proca (GP) theories, its decoupling limit (DL) is not restricted to second order equations of motion while still only propagating three healthy degrees of freedom and coincides with the DL of ghost-free massive gravity in the scalar-vector sector \cite{deRham:2010ik,deRham:2010kj,deRham:2011rn,Ondo:2013wka}. This connection with the decoupling limit of massive gravity naturally raises the question of whether one would expect the UV completion of PN to differ significantly from GP.
The investigation of quantum stability of PN and its quantum comparison to GP is subject of the present work.

The next section \S\ref{PN} is a quick review of the construction of the new effective Proca interactions. By means of a decoupling limit analysis in \S\ref{Dec} we conclude that the EFT is stable under quantum corrections but but find a similar scaling as for the counter-terms arising in GP, suggesting a similar type of UV implementation.
The explicit one-loop effective action in unitary gauge is presented in \S\ref{Feyn} up to quartic order in the fields and confirm our decoupling limit analysis. Both theories have the same quantum implications and the arising counter terms have the same operator structure and the mass scaling.


\section{Proca-Nuevo}\label{PN}

The construction of the PN interactions is heavily inspired by massive gravity and starts by considering the tensorial combination
\begin{equation}
    f_{\mu\nu}[A]=\eta_{\mu\nu} + 2 \frac{\partial_{(\mu}A_{\nu)}}{\Lambda^{2}_{2}}+\frac{\partial_{\mu} A_{\alpha} \partial_{\nu} A_{\beta} \eta^{\alpha\beta}}{\Lambda^{4}_{2}}\,,
\end{equation}
where $\eta_{\mu\nu}$ are the components of the background Minkowski metric\footnote{As opposed to \cite{ProcaNeuvo} we are working with a mostly minus signature $\eta_{\mu\nu}=\text{diag}(+1,-1,-1,-1)$}. This tensor arises in massive gravity \cite{deRham:2010ik} as the result of the covariantization of the spin-2 field through the introduction of four St\"uckelberg fields $\phi^{i} = x^i +\frac{A^i}{\Lambda^2_2}$ in the background metric $f_{\mu\nu}=\eta_{ij}\partial^i \phi_\mu \partial^j \phi_\nu$ which is where the vector field enters in the formulation.

The combination
\be
\mathcal{K}_{\mu\nu}[A] = \left(\sqrt{\eta^{-1} f[A]}\right)_{\mu\nu} - \eta_{\mu\nu}\;,\quad\text{where}\quad\left(\sqrt{\eta^{-1} f[A]}\right)_{\mu\alpha}\left(\sqrt{\eta^{-1} f[A]}\right)^\alpha_{\;\;\nu}=f_{\mu\nu}\,,
\ee
which arises in massive gravity in the formulation of the mass term, serves then as a building block for an usual Galileon-like construction \cite{ProcaNeuvo}
\begin{align}\label{LPN}
\mathcal{L}_{\mathcal{K}}[A]&=\Lambda_2^{4} \sum^{4}_{n=0} \alpha_{n}(A^2) \mathcal{L}_{n} [\mathcal{K}[A]] \,,
\end{align}
where
\begin{align}
    \mathcal{L}_{0} [\mathcal{K}[A]]&= \epsilon_{\mu\nu\alpha\beta}\epsilon^{\mu\nu\alpha\beta}\nonumber\\
    \mathcal{L}_{1} [\mathcal{K}[A]]&= \epsilon_{\mu\nu\alpha\beta}\epsilon^{\mu\nu\alpha}_{\;\;\;\;\;\;\gamma}\mathcal{K}^{\beta\gamma}\nonumber\\
    \mathcal{L}_{2} [\mathcal{K}[A]]&= \epsilon_{\mu\nu\alpha\beta}\epsilon^{\mu\nu}_{\;\;\;\;\gamma\delta} \mathcal{K}^{\beta\delta}\mathcal{K}^{\alpha\gamma}\nonumber\\
    \mathcal{L}_{3} [\mathcal{K}[A]]&= \epsilon_{\mu\nu\alpha\beta}\epsilon^{\mu}_{\;\;\sigma\gamma\delta} \mathcal{K}^{\beta\delta}\mathcal{K}^{\alpha\gamma}\mathcal{K}^{\nu\sigma}\nonumber\\
    \mathcal{L}_{4} [\mathcal{K}[A]]&= \epsilon_{\mu\nu\alpha\beta}\epsilon_{\rho\sigma\gamma\delta} \mathcal{K}^{\beta\delta}\mathcal{K}^{\alpha\gamma}\mathcal{K}^{\nu\sigma}\mathcal{K}^{\mu\rho}\,,
\end{align}
where $\alpha_{n}(A^2)$ are arbitrary polynomial functions. Note that the $n=0$ contribution includes a hard mass term $\frac{1}{2}m^2A^2$.

In the strict sense we may think of this as a standard EFT with cutoff $\Lambda$, but the implementation of the Vainshtein mechanism typically requires trusting the theory for some background above that scale and these types of EFT can be seen to enjoy a reorganization of operator whereby a class of operators may be considered to be large. In this sense $\Lambda$ can be considered here as the strong coupling scale and not necessarily the cutoff, which is why it is meaningful to ask ourselves at what scale do quantum corrections typically enter and that scale will serve as setting up the real cutoff of this EFT.

Concerning the analysis of quantum stability, it is instructive to perturbatively expand the Lagrangian \eqref{LPN} in terms of irrelevant operators in the standard EFT sense. Writing $\alpha_n (A^2) = \Tilde{\alpha}_n +\frac{m^2}{\Lambda^4_2}\Tilde{\gamma}A^2+\frac{m^4}{\Lambda^8_2}\Tilde{\lambda}_n A^4+...$ the Lagrangian up to fourth order takes the following form\footnote{Note the sign differences as compared to \cite{ProcaNeuvo} due to the opposite signature convention.}
\small
\begin{IEEEeqnarray}{rCl}
\mathcal{L}^{(2)}_{\kappa}&=&-\frac{1}{4}F^{\mu\nu}F_{\mu\nu}+\frac{1}{2}m^{2}A^{2}\\
\label{PPN1}\mathcal{L}^{(3)}_{\kappa}&=\frac{1}{\Lambda_2^2}&\left\{-\frac{1}{4}(2\Tilde{\alpha}_{2}-3\Tilde{\alpha}_{3})[F^{2}][\partial A]-\frac{1}{4}(1-4\Tilde{\alpha}_{2}+6\Tilde{\alpha}_{3})(F^{2})_{\mu\nu}\partial^{\mu}A^{\nu}+6\Tilde{\gamma}_{1}m^{2}A^{2}[\partial A]\right\}\\
\mathcal{L}^{(4)}_{\kappa}&=\frac{1}{\Lambda_2^4}&\left\{\frac{1}{32}(\Tilde{\alpha_{2}}-3\Tilde{\alpha_{3}}+6\Tilde{\alpha_{4}}) [F^2]^2 +\frac{1}{64}(5- 20\Tilde{\alpha_{2}}-12\Tilde{\alpha_{3}}+168\Tilde{\alpha_{4}})F^{2}_{\mu\nu}F^{2\mu\nu}\nonumber\right.\\
&&\left.+\frac{3}{8}(\Tilde{\alpha_{3}}-4\Tilde{\alpha_{4}})[F^2]\left([\partial A]^2 -\partial_{\alpha} A_{\beta} \partial^{\beta} A^{\alpha}\right)-\frac{1}{8}F^{2}_{\mu\nu}\partial^{\beta} A^{\mu} \partial_{\beta} A^{\nu}\nonumber\right.\\
&&\left.+\left(\frac{1}{2}\Tilde{\alpha_{2}}+\frac{3}{4}\Tilde{\alpha_{3}}-6\Tilde{\alpha_{4}}\right)F^{2\mu\nu}\left(\partial^{\beta} A_{\mu}\partial_{\beta} A_{\nu} - [\partial A]\partial_{\mu} A_{\nu}\right)\nonumber\right.\\
&&\left.+\left(-\frac{1}{8}+\frac{1}{2}\Tilde{\alpha_{2}}-3\Tilde{\alpha_{4}}\right)F^{\mu\nu}F^{\alpha\beta}\partial_{\mu} A_{\alpha}\partial_{\nu} A_{\beta}\nonumber\right.\\
&&\left.-m^{2} A^{2} \left[2\Tilde{\gamma_{2}}[\partial A]^{2} -\left(\frac{3}{2}\Tilde{\gamma_{1}}+\Tilde{\gamma_{2}}\right)\partial_{\mu} A_{\nu} \partial^{\nu} A^{\mu} +\left(\frac{3}{2}\Tilde{\gamma_{1}}-\Tilde{\gamma_{2}}\right)\partial_{\mu} A_{\nu} \partial^{\mu} A^{\nu}\right]\nonumber\right.\\
&&\left.+24\Tilde{\lambda_{0}} m^4 A^4\right\}\,.\label{PPN2}
\end{IEEEeqnarray}
\normalsize

The decoupling limit is in this case defined via the introduction of a scalar St\"uckelberg field $\phi$ through the replacement
\be\label{Stuckelberg}
A_\mu\rightarrow A_\mu+\tfrac{1}{m}\partial_\mu\phi \,,
\ee
where canonical normalization fixes the mass scale. This formulation restores gauge invariance. The decoupling limit is then defined as the smooth massless limit
\be\label{DL}
m\rightarrow 0 \quad\text{and}\quad \Lambda_2\rightarrow \infty\,,\;\;\text{while}\quad \Lambda_3\equiv\left(\Lambda_2^2m\right)^{\frac{1}{3}}=\text{const}\,,
\ee
where $\Lambda_3$ is the lowest strong coupling scale. For concreteness, \eqref{PPN1} and \eqref{PPN2} the DL read
\small
\begin{IEEEeqnarray}{rCl}
    \mathcal{L}^{(3)}_{DL}&=\frac{1}{\Lambda_3^3}&\left\{-\frac{1}{4}(2\Tilde{\alpha_2}-3\Tilde{\alpha_3})[F^2]\Box \phi - \frac{1}{4} (1-4\Tilde{\alpha_2}+6\Tilde{\alpha_3})F^2_{\mu\nu} \partial^\mu\partial^\nu\phi +6\Tilde{\gamma_1}\partial_\mu\phi\partial^\mu\phi \Box \phi \right\}\\
    \mathcal{L}^{(4)}_{DL}&=\frac{1}{\Lambda_3^6}&\left\{\frac{3}{8} (\Tilde{\alpha_3}-4\Tilde{\alpha_4})[F^2]((\Box \phi)^2-(\partial_\mu \partial_\nu \phi)^2)-\frac{1}{8}F^{2}_{\mu\nu}\partial^\alpha \partial^\mu \phi \partial_\alpha \partial^\nu \phi\nonumber \right.\\
   && \left.+ \left(\frac{1}{2}\Tilde{\alpha_2} +\frac{3}{4}\Tilde{\alpha_3}-6\Tilde{\alpha_4}\right)F^{2\mu\nu}
   \left(\partial^\alpha \partial_\mu \phi \partial_\alpha\partial_\nu \phi -(\Box\phi)\partial_\mu\partial_\nu\phi\right)\nonumber\right.\\
   && \left.+\left(-\frac{1}{8}+\frac{1}{2}\Tilde{\alpha_{2}}-3\Tilde{\alpha_{4}}\right)F^{\mu\nu}F^{\alpha\beta}\partial_{\mu} \partial_{\alpha} \phi\partial_{\nu} \partial_{\beta}\phi\nonumber\right.\\
   && \left.-2 \Tilde{\gamma_2}(\partial_\mu\phi)^2 \left((\Box\phi)^2-(\partial_\mu\partial_\nu\phi)^2\right)\phantom{\frac{1}{2}}\right\}
\end{IEEEeqnarray}
\normalsize

It was proven in \cite{ProcaNeuvo} that this theory is genuinely distinct from Generalized Proca theories, with a constraint structure allowing for higher order equations of motion in the scalar-vector sector while still only propagating three healthy degrees of freedom. However, the pure scalar sector is restricted to a Galileon form.


\section{Quantum Stability and its Decoupling Limit Analysis}\label{Dec}

The decoupling limit is a powerful tool acting as a high energy limit\footnote{In the sense that one considers energies much larger than the vector mass which in the DL translates into $m\ll 1$. A clear separation between scales $m\ll \Lambda_3\ll\Lambda_2$ is therefore crucial and has to be preserved by quantum corrections.} focusing on operators which dominate at the lowest strong coupling scale $\Lambda_3$. Quantum stability translates into the existence of a regime below $\Lambda_3$ for which classical higher order interactions become large while radiative corrections remain negligible. In this limit, the helicity-0 Goldstone mode $\phi$ decouples in a symmetry sense and the theory is approximately described by a massless scalar-vector sector.

Throughout this section we will only be interested in powercounting and will thus consider only the schematic form of terms. In this language, the Lagrangian of PN is built out of three building blocks dictated by dimensional analysis and Lorentz invariance
\be\label{SchematicL}
\mathcal{L}\sim \Lambda_2^4\;\left(\frac{mA}{\Lambda_2^2}\right)^{2a_1}\left(\frac{F}{\Lambda_2^2}\right)^{a_2}\left(\frac{\pd A}{\Lambda_2^2}\right)^{a_3}\,,\quad 0\leq a_{1,2,3}\,,
\ee
where the first term represents the expansion of $\alpha_n (A^2)$ while the others cover the contributions from $\mathcal{K}_{\mu\nu}[A]$. Note that the Lagrangians \eqref{PPN1} and \eqref{PPN2} explicitly rule out the cases for which $a_{1}=0$ and $a_{2}\leq 1$ simultaneously such that we always have either $a_{1}\geq 1$ or $a_{2}\geq 2$. However, this is completely general and applies to all orders: Terms with $a_{2}=0$ which do not reduce to pure scalar Galileons in the DL are total derivatives, while the mixed sector requires an even number of field strength factors. This specific structure is reminiscent of ghost-free massive gravity.

More precisely, the schematic form of the Lagrangian is therefore
\be\label{SchematicL2}
\mathcal{L}\sim m^2A^2\;\left(\frac{mA}{\Lambda_2^2}\right)^{2a_1}\left(\frac{F}{\Lambda_2^2}\right)^{a_2}\left(\frac{\pd A}{\Lambda_2^2}\right)^{a_3}+\; F^2\;\left(\frac{mA}{\Lambda_2^2}\right)^{2b_1}\left(\frac{F}{\Lambda_2^2}\right)^{b_2}\left(\frac{\pd A}{\Lambda_2^2}\right)^{b_3}\, .
\ee
This is crucial, as otherwise the decoupling limit would diverge and cease to be well defined. The general structure of the Lagrangian \eqref{SchematicL2} exactly matches the one found in the case of Generalized Proca theories \cite{Heisenberg:2020jtr} such that the same analysis of quantum stability essentially goes through. 
It is worthwhile to emphasize that if one had a more generic Proca theory where the DL took a different form, the nature of the quantum corrections would differ and the cutoff of the theory would be significantly lower. This property is at the heart of why we may be bothered into these kinds of EFTs in the first place as opposed to something more generic where all operators are thrown in at the same scale.

In brief\footnote{We refer to \cite{Heisenberg:2020jtr} for details.},
the St\"uckelberg trick allows to take the decoupling limit \eqref{DL} on each individual operator. We obtain that indeed the limit is well defined and the classical Lagrangian reduces to
\be\label{SchematicLDL}
\mathcal{L}_{\text{DL}}\sim (\pd\phi)^2\;\left(\frac{\pd^2\phi}{\Lambda_3^3}\right)^{a_3}+\;F^2 \;\left(\frac{\pd^2\phi}{\Lambda_3^3}\right)^{b_3}\,,
\ee
where now $0\leq a_{3}\leq 3$ as all the $a_{3}\geq4$ terms are total derivatives and have vanishing equations of motion \cite{Heisenberg:2014rta,Jimenez:2016isa,Heisenberg:2018vsk,Jimenez:2019hpl}. This nicely reflects the fact that for the terms involving only the scalar field $\phi$ the individual classical operators only remain ghost-free in the specific scalar Galileon form.

At one loop, the specific form of \eqref{SchematicLDL} implies that each vertex in the decoupling limit comes at least with a factor of $1/\Lambda_3^3$. This means that in dimensional regularization and only considering $1$PI diagrams at one loop, where each vertex at least includes one external leg while two legs are contributing to the loop, there are only two distinct schematic building blocks for quantum induced operators \small$\frac{\pd F}{\Lambda_3^3}\;$\normalsize and \small$\frac{\pd^2\phi}{\Lambda_3^3}\;$\normalsize. Therefore, a general one loop counterterm in the decoupling limit has the generic form\footnote{This schematic form is fixed through Lorentz invariance, power-counting and the well behaved propagators in the decoupling limit.}
\be\label{fullDL}
\mathcal{L}^{\text{c}}_{\text{DL}}\sim \pd^4\left(\frac{\pd F}{\Lambda_3^{3}}\right)^{2 c_2}\left(\frac{\pd^2\phi}{\Lambda_3^{3}}\right)^{c_3}\sim
\begin{cases}
F^2\left(\frac{\pd^2}{\Lambda_3^2}\right)^{2+c_2}\left(\frac{F^2}{\Lambda_3^4}\right)^{c_2-1}\left(\frac{\pd^2\phi}{\Lambda_3^3}\right)^{c_3}&,\quad c_2\geq 1\\
\\
(\pd\phi)^2\left(\frac{\pd^2}{\Lambda_3^2}\right)^{3+c_2}\left(\frac{F^2}{\Lambda_3^4}\right)^{c_2}\left(\frac{\pd^2\phi}{\Lambda_3^3}\right)^{c_3-2}&,\quad c_3\geq 2
\end{cases}
\ee
where $2 c_2+c_3=N\geq 2$ with $N$ the number of external fields and $c_{2,3}\geq0$ positive integers.\footnote{The two possible reformulations in \eqref{fullDL} result from comparing the operator to the two kinetic terms $F^2$ and $(\pd\phi)^2$ of the theory. For most $b_i$ values either one can be employed. Only for $b_1=0$ the upper one looses its sense, while the lower one is not valid whenever $b_3<2$.} We can therefore identify one classical and two quantum expansion parameters
\be\label{Parameters}
\alpha_{\text{cl}}=\frac{\partial^2\phi}{\Lambda_3^3}\;,\quad \alpha_{\text{q}}=\frac{\partial^2}{\Lambda_3^2}\;,\quad\alpha_{\tilde{\text{q}}}=\frac{F^2}{\Lambda_3^4}\,,
\ee
and write the total EFT Lagrangian as an expansion in these parameters.

One can even generalize this analysis to higher loops. Each additional loop comes with an increase in factors of $1/\Lambda_3^3$ compared to the same diagram without the additional loop. This is because in order to add a loop to a diagram while keeping the number of external legs fixed necessarily requires the inclusion of an additional vertex or the addition of legs to existing vertices. Now since the number of external legs remains the same in this comparison, to match dimensions these factors can only be compensated with additional powers of derivatives. Thus, higher loops will merely introduce additional factors of $\alpha_{\text{q}}$.\footnote{Actually, Lorentz invariance requires the additional factor to be $\pd^6/\Lambda_3^6$.}

From here on, the analysis exactly parallels the one employed for the consolidation of radiative stability of various derivative self-interacting theories such as scalar Galileons \cite{Luty:2003vm,Nicolis:2004qq,Burgess:2006bm,Hinterbichler:2010xn,Hinterbichler:2011tt,deRham:2012ew,Goon:2016ihr,Heisenberg:2020cyi,Heisenberg:2020jtr}: The complete EFT Lagrangian can be written as an expansion in the three parameters $\alpha_{\text{cl}}$, $\alpha_{\text{q}}$ and $\alpha_{\tilde{\text{q}}}$
\be\label{CQS}
\mathcal{L}_{\text{DL}}\sim\left(F^2+(\pd\phi)^2\right)\,\alpha_{\text{cl}}^{b_3}+\left(F^2+(\pd\phi)^2\right)\,\alpha_{\text{q}}^{2+n}\alpha_{\tilde{\text{q}}}^l\,\alpha_{\text{cl}}^m\,,\quad 0\leq b_3,l,n,m\,,
\ee
where only the quantum induced operators carry the quantum parameters $\alpha_{\text{q},\tilde{\text{q}}}$. More precisely, quantum corrections always involve at least two powers of $\alpha_{\text{q}}$ compared to the initial PN Lagrangian which marks a clear separation between classical and quantum terms and implies non-renormalization of classical terms within the DL.\footnote{Non-renormalization in the weak sense, tied to dimensional regularization (see \cite{Goon:2016ihr}).} In specific applications, we therefore expect the existence of a regime below the energy scale $\Lambda_3$ where quantum contributions are heavily suppressed $\alpha_{\text{q},\tilde{\text{q}}}\ll 1$, while classical non-linear terms, although equally non-renormalizable, are important compared to the kinetic term $\alpha_{\text{cl}}\sim\mathcal{O}(1)$.

One could be worried about the regime $\alpha_{\text{cl}}\gg 1$ in the expansion in external legs. Naively we would need to conclude, that the EFT breaks down as terms with higher derivatives and higher numbers of background fields are considered, regardless of whether $\alpha_{\text{q},\tilde{\text{q}}}\ll 1$ or not. However, from previous analysis \cite{Nicolis:2004qq,deRham:2013qqa} one would expect that the tree level kinetic term of quantum fluctuations gets enhanced by large classical-non-linearities, hence, precisely in the regime $\alpha_{\text{cl},\tilde{\text{cl}}}\gg 1$. As long as the classical contributions do not lead to ghost instabilities as is the case by construction, the quantum fluctuations are rather further suppressed on such scales in contrast to what one could have expected.

Of course, the clear separation between classical and quantum terms \eqref{CQS} is only valid in the decoupling limit where crucially propagators behave as $\sim\frac{1}{p^2}$ and in the initial, unitary gauge formulation quantum corrections will generate operators with the same form as classical operators, leading to a potential detuning and introduction of ghost instabilities. The beauty of the DL analysis is however, that since it focuses on the relevant scale $\Lambda_3$ we know that all operators which did not survive the decoupling limit are further suppressed by factors of $m/\Lambda_3$ in comparison to operators present in the DL such that they have no impact on the viability of the EFT. For example we expect that loop corrections will introduce a counterterm of the form
\be\label{Odetuning}
\sim\frac{m^4}{\Lambda_2^4}(\pd_\mu A^\mu)^2\,,
\ee
which explicitly detunes the classical gauge invariant kinetic term and thus introduces a dynamical ghostly temporal component $A^t$. This term is however heavily suppressed, which can alternatively be understood as the statement that after canonically normalizing, the mass of the associated ghost $m_t^2\sim\Lambda_3^6/m^4$ lies way above the cutoff of the theory.

Moreover, the commutativity of decoupling limit and quantum correction calculations allows one to infer the form of counterterms in the unitary gauge from the expansion \eqref{fullDL}. Concretely, the least suppressed quantum corrections in the original formulation need to scale like
\be\label{lscont}
\mathcal{L}^{\text{c}}\sim \pd^4\left(\frac{\pd F}{\Lambda_2^2 m}\right)^{2c_2}\left(\frac{\pd A}{\Lambda_2^2}\right)^{c_3}\,.
\ee
Hence, in the limiting case $c_3=0$, quantum corrections need to preserve gauge invariance in order not to spoil the stability properties in the DL discussed in this section. This will be beautifully confirmed by our explicit calculations in the next section.

Since the DL anaysis of quantum corrections exactly parallels the GP case, only explicit calculation will be able to tell whether PN has any improved UV behavior, hence cancellations between terms.


\section{One-loop calculations}\label{Feyn}

In this section we offer explicit calculations of one-loop corrections in the unitary gauge up to four-point at the one-loop level using standard Feynman diagram techniques.

Following a $\overline{\text{MS}}$-scheme, the one-loop counterterms can be inferred from the UV divergence of the $1$PI diagrams which we will compute using dimensional regularization. We are thus after the log-divergent part of the one-loop $1$PI diagrams with $N$ external legs $\mathcal{M}_N^{\text{div}}$ which will be a function of the external momenta $p_i$, ${\scriptstyle i=1,..,N-1}$. Throughout this work we will treat all momenta as incoming.

\subsection{Two-point}

\begin{figure}[H]
\begin{center}
\begin{fmffile}{loops2pf}
\begin{fmfgraph*}(125,65)
     \fmfleft{i}
     \fmfright{o}
     \fmf{plain,tension=3}{i,v1}
     \fmf{plain,left=1}{v1,v2}
     \fmf{plain,left=1}{v2,v1}
     \fmf{plain,tension=3}{v2,o}
     \fmfdot{v1,v2}
    \end{fmfgraph*}
    \qquad \qquad
  \begin{fmfgraph*}(125,65)
 	\fmfleft{i}
     \fmfright{o}
     \fmf{plain,tension=3}{i,v1}
     \fmf{plain}{v1,v1}
     \fmf{plain,tension=3}{v1,o}
     \fmfdot{v1}
\end{fmfgraph*}
\end{fmffile}
\end{center}
\caption{The two possible 1PI two-point diagrams, each having a symmetry factor of $2$.}
\label{loop2pt}
\end{figure}
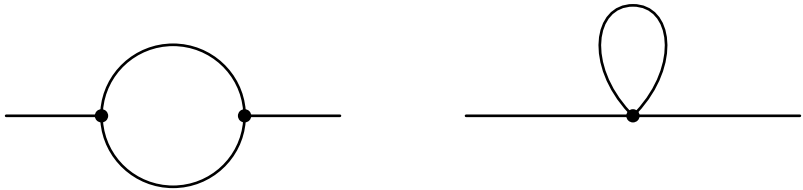

The perturbative renormalization procedure of the two-point function only requires the calculation of two distinct 1PI one loop diagrams depicted in Fig.\ref{loop2pt}. Hence, it is enough to explicitly know the two lowest order interaction terms \eqref{PPN1} and \eqref{PPN2}. Explicit Feynman rules can be found in the appendix \ref{fr}.

Using a dimensional regularization procedure\footnote{Note that at one loop the divergent part is blind to the extra factors of $d$ in the Levi-Civita contractions, such that we will disregard them.} with $d=4+2\epsilon$, the divergent part of the reduced matrix element up to two powers of momenta reads\footnote{As a cross check, we computed the lower order contributions with a complementary effective action based generalized Schwinger-DeWitt method and found perfect agreement. See \cite{Heisenberg:2020jtr} for details regarding this method.}

\small
\begin{IEEEeqnarray}{rCl}\label{2ptFeyn}
\mathcal{M}_2^{\text{div}}&= \frac{\epsilon_1^\mu \epsilon_2^\nu}{3072 \epsilon \Lambda_2^6 \pi^2}&\Big[ 576 m^6 g_{\mu\nu}(-3 \tilde{\gamma}_1 +36 \tilde{\gamma}_1^2 +2(\tilde{\gamma}_2 +72\tilde{\lambda}_0))\nonumber\\
&&-6m^4\Big([176 \tilde{\alpha}_2^2 -48\tilde{\alpha}_2 (-2 +11\tilde{\alpha}_3 +64 \tilde{\gamma}_1)+3 (11-52\tilde{\alpha}_3 +132\tilde{\alpha}_3^2 +16 \tilde{\alpha}_4 \nonumber\\
&&-160 \tilde{\gamma}_1 +1536 \tilde{\alpha}_3 \tilde{\gamma}_1 +4608 \tilde{\gamma}_1^2 +64 \tilde{\gamma}_2)]p_\mu p_\nu+ 2 [32\tilde{\alpha}_2^2 -12 \tilde{\alpha}_2 (3+8\tilde{\alpha}_3 -32\tilde{\gamma}_1)\nonumber\\
&& +3 (-1 +20\tilde{\alpha}_3 +24\tilde{\alpha}_3^2 -8\tilde{\alpha}_4 -16\tilde{\gamma}_1 -192 \tilde{\alpha}_3 \tilde{\gamma}_1 -576 \tilde{\gamma}_1^2 -32 \tilde{\gamma}_2)]p.p g_{\mu\nu}\Big)\nonumber\\
&\hspace{1.5cm}+2m^2 p^2 &\Big(-4[7+16\tilde{\alpha}_2^2 +84\tilde{\alpha}_3 +36\tilde{\alpha}_3^2 -8\tilde{\alpha}_2 (7+6\tilde{\alpha}_3-36\tilde{\gamma}_1 )+144 \tilde{\gamma}_1 -432 \tilde{\alpha}_3 \tilde{\gamma}_1\nonumber\\
&& -2592 \tilde{\gamma}_1^2]p_\mu p_\nu+[31+304\tilde{\alpha}_2^2 +228 \tilde{\alpha}_3 +684\tilde{\alpha}_3^2 -152\tilde{\alpha}_2 (1+6\tilde{\alpha}_3)]p^2 g_{\mu\nu}\Big)\nonumber\\
&\hspace{2cm}-p^4&\Big([-13-208\tilde{\alpha}_3^2 -228\tilde{\alpha}_3 -468\tilde{\alpha}_3^2 +8\tilde{\alpha}_2 (19+78\tilde{\alpha}_3)+3456\tilde{\gamma}_1^2]p_\mu p_\nu\nonumber\\
&&+16(1-4\tilde{\alpha}_2 +6\tilde{\alpha}_3)^2 p^2 g_{\mu\nu}\Big)\nonumber\\
&&-\frac{2(1-4\tilde{\alpha}_2 +6\tilde{\alpha}_3)^2 p^6}{m^2}\Big(p_\mu p_\nu -p^2 g_{\mu\nu}\Big)\Big]
\end{IEEEeqnarray}

\normalsize
This result confirms the one-loop quantum stability of the theory at two point as all the contributions remain suppressed enough in order to preserve the classical structure below the cutoff $\Lambda_3$. In particular no $O(\frac{1}{\Lambda_2^4 m^4})$ contribution is present which could not have been excluded through naive powercounting in the unitary gauge. Moreover, as predicted by the decoupling limit analysis \S\ref{Dec} in the limiting case $O(\frac{1}{\Lambda_2^4 m^2})$ corresponding to $c_3=0$, $c_2=1$ in \eqref{lscont} the induced counter terms precisely acquire a gauge invariant structure such that, now taking the decoupling limit after renormalizing, the counterterms fit into the expansion \eqref{fullDL}
\begin{equation}
     \frac{\partial^{6}}{\Lambda_2^4 m^2}F^2 \xrightarrow{DL} \frac{\partial^{6}}{\Lambda_3^6}F^2\,.
\end{equation}

However, at this level we do not find any improved behavior as compared to Generalized Proca theories.

\subsection{Three-point}

\begin{figure}[H]
\begin{center}
\begin{fmffile}{loops3pf}
\begin{fmfgraph*}(125,90)
     \fmfleft{i1,i2}
     \fmfright{o}
     \fmf{plain,tension=3}{i1,v1}
     \fmf{plain,tension=3}{i2,v2}
     \fmf{plain}{v1,v2}
     \fmf{plain}{v1,v3}
     \fmf{plain}{v2,v3}
     \fmf{plain,tension=3}{v3,o}
     \fmfdot{v1,v2,v3}
    \end{fmfgraph*}
    \qquad
  \begin{fmfgraph*}(125,90)
 	\fmfleft{i1,i2}
     \fmfright{o}
      \fmf{plain,tension=3}{i1,v1}
     \fmf{plain,tension=3}{i2,v1}
     \fmf{plain,left=1}{v1,v2}
     \fmf{plain,left=1}{v2,v1}
     \fmf{plain,tension=3}{v2,o}
     \fmfdot{v1,v2}
\end{fmfgraph*}
\qquad
 \begin{fmfgraph*}(125,90)
 	\fmfleft{i1,i2}
     \fmfright{o}
     \fmf{plain,tension=3}{i1,v1}
     \fmf{plain,tension=3}{i2,v1}
     \fmf{plain}{v1,v1}
     \fmf{plain,tension=3}{v1,o}
     \fmfdot{v1}
\end{fmfgraph*}
\end{fmffile}
\end{center}
\caption{\small{Three distinct one-loop $1$PI diagrams giving rise to corrections of the three point function. The first one arising from three $\mathcal{L}_3$ vertex contribution has a symmetry factor of 1, while the other two come with a symmetry factor of 2. The second diagram combining $\mathcal{L}_3$ and $\mathcal{L}_4$ has multiplicity 3.}}
\label{loop3pf}
\end{figure}
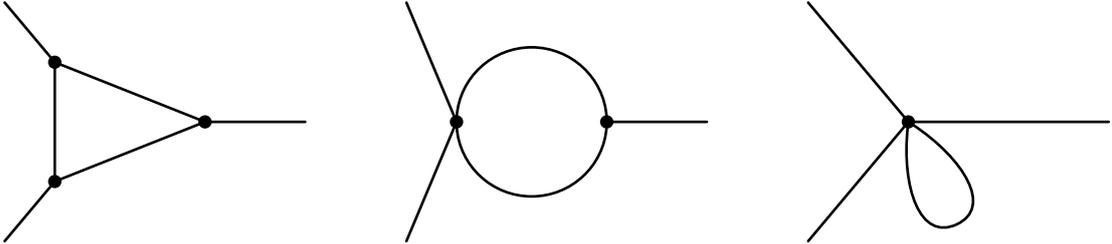

At three-point, three 1PI one-loop diagrams contribute (see fig.\ref{loop3pf}). Note that the third one is a contribution from $\mathcal{L}_5$ with associated Feynman rule which we do not show explicitly in the main text, but can be found in the auxiliary file.

We are mainly interested in the scaling of terms present, which is why we will present the rather lengthy results in a schematic sum of contributions $M^{\scaleto{(i)\mathstrut}{6pt}}_{3}$ where $i$ denotes the power of external momenta involved
\begin{align}
    \mathcal{M}_3^{\text{div}}&= \frac{1}{\Lambda^6_2}\left[m^6 \mathcal{M}^{(1)}_{(34)} +m^4 \mathcal{M}^{(3)}_{(34)} + m^2 \mathcal{M}^{(5)}_{(34)} + \mathcal{M}^{(7)}_{(34)} + \frac{1}{m^2} \mathcal{M}^{(9)}_{(34)}\right]\,.
\end{align}
Again, as expected from the decoupling limit analysis, there is no $\sim k^{11}$ present which could destabilize the EFT. However, at three-point as well we do not observe any cancellation between terms which would signal an improved UV behavior of PN compared with Generalized Proca. Since neither at 2-point nor at 3-point we found any particular improved behavior of one-loop corrections, we conclude that PN as an EFT is stable under quantum corrections and behaves exactly like GP in that respect.


\section{Conclusion}
GR generalisations to incorporate various cosmological issues requires inclusion of additional degrees of freedom. As has been pointed out in several literature that, no linear terms addition would satisfy both, theoretical consistency and observational equivalence with GR at large length scales, one naturally moves towards non-linear terms additions. We saw that the procedure to include these extra non-linear terms does not necessarily force us to look at interaction terms which result in at most second order equations of motion (like in GP), rather one can also have higher order terms present in the theory which can still propagate the required degrees (as shown in PN). In PN all the non-linear terms only appear relevant above a certain energy scale (below a certain distance scale, which happened to be much smaller than the Planck scale) and were highly suppressed below energy $\Lambda_3$. This would mean that all the large distance scale phenomenon can be safely reduced to GR (as observed).

Although ghostly degrees can be removed from the classical picture, they reappear after quantum corrections. It was not their presence which was bothering, but rather the scale at which they become relevant ($\Lambda_3$, in our case). Now, whereas scalar mode additions did produce some interesting results in the past, vector field generalisations seem to be producing much more varied and interesting physics (such as higher cut-off). As massless vector fields generalisation suffered the no-go obstacle, we resorted to the study of the massive ones (specially with the higher order derivative interactions). Although early studies indicated a generalisation of Proca action to a theory (GP) with 5 uniquely identified higher order derivative interactions, a recent study which benefited from an inherent non-linear mechanism in massive gravity derivation produced the Proca-Neuvo action with exact resemblance with the massive gravity Decoupling limit (owing to its derivation of course). In this work, we proved the Quantum stability of the Proca-Neuvo theory below a specific UV cut-off ($\Lambda_3$). Moreover, although PN and GP are different inherently in their classical structure (no resemblance either in field redefinition or even at the tree level), both have similar high energy behaviour when quantum corrections are taken into account. This can be directly checked by visual inspection and comparison with \cite{Heisenberg:2020jtr}. A possible conclusion would be that they both have their origin at a single higher UV complete theory which we leave for future studies. An important part which is still left in this work is the complete covariantization of the PN action on a perturbed Minkowski background (beyond its massive gravity covariantization).
\label{discussion}


\section*{Acknowledgments}
LH is supported by funding from the European Research Council (ERC) under the European Unions Horizon 2020 research and innovation programme grant agreement No 801781 and by the Swiss National Science Foundation grant 179740.

\newpage


\appendix

\section{Feynman Rules}\label{fr}

In our signature convention, the Feynman propagator is
\be
i\Delta^{\mu\nu}(p) = \frac{i}{p^2 - m^2 +i\epsilon}\left(-g^{\mu\nu} + \frac{p^\mu p^\nu}{m^2}\right)
\ee
while the 3 and 4-leg vertecies for all incoming momenta read
\small
\begin{IEEEeqnarray}{rCl}\label{finalResultsG2}
-i\Lambda_2^2 V_3&=&-12 \Tilde{\gamma_1} m^2 p_1^{\mu_1} g^{\mu_2\mu_3}+\left(3 \Tilde{ \alpha_3}-2 \Tilde{ \alpha_2}\right) p_1^{\mu_1} p_2.p_3 g^{\mu_2\mu_3} +\left(\frac{2}{3} \Tilde{ \alpha_2}- \Tilde{ \alpha_3}\right) p_1^{\mu_1} p_2^{\mu_3} p_3^{\mu_2}\nonumber\\
  && +\left( \Tilde{\alpha_2}-\frac{3 \Tilde{ \alpha_3}}{2}-\frac{1}{4}\right) \left[p_1^{\mu_2} p_2.p_3 g^{\mu_1\mu_3}+p_1^{\mu_3} p_2.p_3
   g^{\mu_1\mu_2}-p_1^{\mu_2} p_2^{\mu_3} p_3^{\mu_1}-p_1^{\mu_3} p_2^{\mu_1} p_3^{\mu_2}\right]\\
 &&+\text{cycl}\{1,2,3\} \nonumber
\end{IEEEeqnarray}
\begin{IEEEeqnarray}{rCl}\label{finalResultsG3}
   \Lambda_2^4 V_4 &=&(p_1.p_2) (p_3.p_4)\left[\frac{1}{32} (-4 \Tilde{\alpha}_2+12 \Tilde{\alpha}_3-24 \Tilde{\alpha}_4+1) \eta^{\mu_{3}\mu_{2}} \eta^{\mu_{4}\mu_{1}}+\frac{1}{8} (\Tilde{\alpha}_2-3 \Tilde{\alpha}_3+6 \Tilde{\alpha}_4) \eta^{\mu_{3}\mu_{1}} \eta^{\mu_{4}\mu_{2}}\right]\nonumber\\
   &&+\left(-\Tilde{\alpha}_2-\frac{3 \Tilde{\alpha}_3}{2}+12 \Tilde{\alpha}_4\right)p_1^{\mu_{1}} p_2^{\mu_{4}}(p_3.p_4) \eta^{\mu_{3}\mu_{2}}+\frac{1}{64}(12 \Tilde{\alpha}_2-12 \Tilde{\alpha}_3-24 \Tilde{\alpha}_4-11) p_1^{\mu_{2}} p_2^{\mu_{4}} (p_3.p_4) \eta^{\mu_{3}\mu_{1}}\nonumber\\
   &&-\frac{3}{16} (p_3.p_4) p_1^{\mu_{3}} p_2^{\mu_{1}} (\Tilde{\alpha}_2+\Tilde{\alpha}_3-10 \Tilde{\alpha}_4) \eta^{\mu_{4}\mu_{2}}+(p_2.p_3) p_1^{\mu_{1}} p_4^{\mu_{3}} \left(\frac{\Tilde{\alpha}_2}{2}+\frac{3 \Tilde{\alpha}_3}{4}-6 \Tilde{\alpha}_4\right) \eta^{\mu_{4}\mu_{2}}\nonumber\\
    &&+\frac{1}{64} (p_2.p_3) p_1^{\mu_{2}} p_4^{\mu_{3}} (28 \Tilde{\alpha}_2+36 \Tilde{\alpha}_3-312 \Tilde{\alpha}_4+1) \eta^{\mu_{4}\mu_{1}}+(p_1.p_2) p_3^{\mu_{1}} p_4^{\mu_{3}} \left(\frac{\alpha_2}{2}+\frac{3 \Tilde{\alpha}_3}{4}-6 \Tilde{\alpha}_4\right) \eta^{\mu_{4}\mu_{2}}\nonumber\\
    &&+\frac{3}{64} (p_2.p_3) p_1^{\mu_{3}} p_4^{\mu_{4}} (4 \Tilde{\alpha}_2-4 \Tilde{\alpha}_3-8 \Tilde{\alpha}_4-1) \eta^{\mu_{1}\mu_{2}}+\frac{1}{64} (p_1.p_2) p_3^{\mu_{1}} p_4^{\mu_{4}} (-12 \Tilde{\alpha}_2-36 \Tilde{\alpha}_3+216 \Tilde{\alpha}_4-5) \eta^{\mu_{3}\mu_{2}}\nonumber\\
    &&+\frac{1}{32} (p_2.p_3) p_1^{\mu_{3}} p_4^{\mu_{1}} (4 \Tilde{\alpha}_2+12 \Tilde{\alpha}_3-72 \Tilde{\alpha}_4-1) \eta^{\mu_{4}\mu_{2}}+\frac{1}{16} (p_2.p_3) p_1^{\mu_{2}} p_4^{\mu_{1}} (-\Tilde{\alpha}_2+3 \Tilde{\alpha}_3-6 \Tilde{\alpha}_4) \eta^{\mu_{3}\mu_{4}}\nonumber\\
    &&+\frac{1}{32} (p_1.p_2) p_3^{\mu_{3}} p_4^{\mu_{1}} (-20 \Tilde{\alpha}_2-12 \Tilde{\alpha}_3+168 \Tilde{\alpha}_4+5) \eta^{\mu_{4}\mu_{2}}+\frac{1}{32} (p_1.p_2) p_3^{\mu_{2}} p_4^{\mu_{1}} (4 \Tilde{\alpha}_2-12 \Tilde{\alpha}_3+24 \Tilde{\alpha}_4-1) \eta^{\mu_{3}\mu_{4}}\nonumber\\
    &&+\frac{3}{4} (p_3.p_4) (\Tilde{\alpha}_3-4 \Tilde{\alpha}_4) p_1^{\mu_{1}} p_2^{\mu_{3}} \eta^{\mu_{4}\mu_{2}}+(p_3.p_4) \left(\Tilde{\gamma}_2 m^2-\frac{3 \Tilde{\gamma}_1 m^2}{2}\right) \eta^{\mu_{3}\mu_{1}} \eta^{\mu_{4}\mu_{2}}+\frac{1}{2} m^2 (3 \Tilde{\gamma}_1+2 \Tilde{\gamma}_2) p_3^{\mu_{2}} p_4^{\mu_{4}} \eta^{\mu_{3}\mu_{1}}\nonumber\\
    &&-2 \Tilde{\gamma}_2 m^2 p_3^{\mu_{4}} p_4^{\mu_{2}} \eta^{\mu_{3}\mu_{1}}+24 \Tilde{\lambda}_0 m^4 \eta^{\mu_{3}\mu_{1}} \eta^{\mu_{4}\mu_{2}}+p_1^{\mu_{1}} p_2^{\mu_{4}} p_3^{\mu_{2}} p_4^{\mu_{3}} \left(\frac{\Tilde{\alpha}_2}{2}+\frac{3 \Tilde{\alpha}_3}{4}-6 \Tilde{\alpha}_4\right)\nonumber\\
    &&+\frac{1}{64} (-20 \Tilde{\alpha}_2-12 \Tilde{\alpha}_3+168 \Tilde{\alpha}_4+5)p_1^{\mu_{4}} p_2^{\mu_{1}} p_3^{\mu_{2}} p_4^{\mu_{3}} +\frac{1}{16} (\Tilde{\alpha}_2+3 (\Tilde{\alpha}_3-6 \Tilde{\alpha}_4))p_1^{\mu_{3}} p_2^{\mu_{1}} p_3^{\mu_{2}} p_4^{\mu_{4}} \nonumber\\
    &&+\frac{1}{16}(\Tilde{\alpha}_2+3 (\Tilde{\alpha}_3-6 \Tilde{\alpha}_4)) p_1^{\mu_{2}} p_2^{\mu_{4}} p_3^{\mu_{3}} p_4^{\mu_{1}} +\frac{1}{64}(-20 \Tilde{\alpha}_2-12 \Tilde{\alpha}_3+168 \Tilde{\alpha}_4+5) p_1^{\mu_{3}} p_2^{\mu_{4}} p_3^{\mu_{2}} p_4^{\mu_{1}} \nonumber\\
    &&+\left(3 \Tilde{\alpha}_4-\frac{3 \Tilde{\alpha}_3}{4}\right) p_1^{\mu_{1}} p_2^{\mu_{3}} p_3^{\mu_{2}} p_4^{\mu_{4}}+\text{cycl}\{1,2,3,4\}
\end{IEEEeqnarray}

\normalsize

\newpage
\bibliographystyle{JHEP}
\bibliography{references}
\end{document}